# Quantifying the effect of oxygen on micro-mechanical properties of a near-alpha titanium alloy

H. M. Gardner[1], P. Gopon[2], C. M. Magazzeni[1], A. Radecka[3], K. Fox[3], D. Rugg[3], J. Wade[2], D. E. J. Armstrong[1], M. P. Moody[1] and P. A. J. Bagot[1]

1. Department of Materials, University of Oxford, Oxford OX1 3PH, United Kingdom.

2. Department of Earth Sciences, University of Oxford, Oxford OX1 3AN, United Kingdom.

3. Rolls-Royce plc, Derby, DE24 8BJ, United Kingdom.

Corresponding Author: hazel.gardner@materials.ox.ac.uk

## Abstract

Atom probe tomography (APT), electron probe microanalysis (EPMA) and nanoindentation were used to characterise the *oxygen-rich layer* on an *in-service* jet engine compressor disc, manufactured from the titanium alloy TIMETAL 834. Oxygen ingress was quantified and related to changes in mechanical properties through nanoindentation studies. The relationship between oxygen concentration, microstructure, crystal orientation and hardness has been explored through correlative hardness mapping, EPMA and electron backscatter diffraction (EBSD). The role of microstructure on oxygen ingress has been studied and oxygen ingress along a potential α/ β interface was directly observed on the nanoscale using APT.

## 1. Introduction

Titanium alloys have been incorporated into jet engine design since the early 1950s due to their corrosion resistance and high specific fatigue strength, which helps minimise total engine weight. In turn, this improves fuel efficiency, providing cost savings to the operator, as well as reducing $CO_2$ emissions. Titanium alloys are used for multiple safety-critical jet engine components including fan blades and compressor discs. For example, the near-α titanium alloy TIMETAL 834, which contains both the hexagonal α phase and cubic β phase, is used to make compressor discs. However, greater use of titanium alloys in compressor discs is currently temperature-limited since above ~480°C, oxygen can start to diffuse through the surface oxide layer on components [1]. This leads to the detrimental formation of a brittle, high oxygen content surface layer, which limits use by promoting crack formation when components are operated at temperatures approaching 600°C under cyclic loading conditions over thousands of hours [2,3].

Historically there has been some ambiguity in the language used to describe high oxygen content layers formed on titanium components. Satko *et al.* [4] provide a helpful distinction between the *oxygen-rich layer* that forms at intermediate *in-service* temperatures (up to ~600°C), and *α case*, which they define as forming at the higher temperatures experienced during alloy processing (~850-1000°C). While both *α case* and *oxygen-rich layer* formation can affect mechanical properties, formation of the *oxygen-rich layer* does not result in microstructural change. In contrast, *α case* formation results in *β denudation* where the volume fraction of the β phase is reduced due to oxygen stabilising the hexagonal α phase. The term '*α case*' is often broadly used in industry as an umbrella term for both phenomena described by Satko *et al.* [4]. However, in this study, the specific definitions of *oxygen-rich layer* and *α case,* as defined above, will be used.

*In-service* formation of an *oxygen-rich layer* degrades the fatigue properties of the component, reducing service life and producing potential safety-critical issues. Accurate component life predictions are required to both improve jet engine efficiency and prevent *in-service* component failures. Such service life predictions rely on a full understanding of both the oxygen ingress depth and the effect of the *oxygen-rich layer* on mechanical properties under *in-service* conditions.

Oxygen ingress depth can be estimated by modelling diffusion of oxygen and nitrogen through application of Fick's second law of diffusion, calibrated with experimental data, to estimate a diffusion coefficient. Multiple attempts have been made at this [1,5,6], but the effect of crystal orientation and microstructure is rarely accounted for, resulting in a large discrepancy in the reported diffusion coefficients. For example, McReynolds *et al.* [5] estimated the oxygen diffusion coefficient in a Ti-6242S sample with a bimodal microstructure to be ~4 X $10^{-3}$ mm$^2$/s at 649°C, which is ~20 times higher than the value reported by Shamblen *et al.* [6] at 639°C.

A variety of models are available to predict the effect of oxygen ingress depth on mechanical properties [7–9]. Parthasarathy *et al.* [8] predict the tendency for surface crack formation in a brittle oxygen-rich surface layer on Ti-6242 as a function of time-temperature-environment history. The Parthasarathy model requires accurate prediction of oxygen ingress depth, but the oxygen ingress was calibrated using the diffusion coefficient of oxygen in the β phase, despite the alloy in the model only containing ~10% volume fraction of β. This demonstrates the need for accurate diffusion coefficients, which account for microstructure and crystal orientation, in order to predict component lifetimes.

Previous studies have also underlined the importance of accurate quantification of oxygen concentration at the surface of titanium [10,11]. Attempts have also been made to quantify oxygen using Secondary Ion Mass Spectroscopy (SIMS) [12]. This is challenging as the interpretation of any SIMS data relies on comparison with suitable micro-analytical standards of similar composition and microstructure to the unknown. Unfortunately, it is non-trivial to find representative standards with independently verified oxygen contents spanning the range exhibited by test samples. An alternative technique is Atom Probe Tomography (APT), a 3D microscopy technique with a uniquely powerful combination of very high spatial and chemical resolution. APT does not rely on external standards for quantification, unlike SIMS, and this enables use of APT to quantify and map oxygen segregation on the nanoscale, thereby generating new insights into the interaction of microstructure and local oxygen concentration. For example, APT studies have revealed formation of $\alpha_2$ precipitates of varying morphology within the *α case* layer formed on Ti-6Al-4V [13] and have also investigated the relationship between mechanical properties and oxygen concentration in this alloy [14]. APT has also been used to study ordering [15–17], oxygen and carbon segregation [18] and fine scale precipitation [19,20] in a range of titanium alloys.

While accurate oxygen quantification is important, it is vital to be able to link changes in oxygen content to mechanical properties. Liu and Welsch [21] determined an empirical relationship between oxygen content and Vickers hardness, shown in Equation 1. The relationship was calculated from bulk micro hardness measurements on a set of Ti-6Al-2V alloys with varying oxygen concentration (0.2-1.83 at. % O).

$$H = H_T + b[O]^{0.5} \qquad \text{(Equation 1)}$$

where $H_T$ is the hardness of an oxygen free alloy, obtained by extrapolation of hardness data to zero oxygen concentration, *b* is a constant and *[O]* is oxygen concentration in weight percent. Building on this, numerous indentation profiles have been measured at the surface of titanium alloys exposed to oxygen [4,14,22,23]. For example, nanoindentation profiles at the surface of Ti-64 heated in air showed the change in hardness as a function of depth from the surface, and found that the hardness profile was steeper for the sample exposed to air at higher partial pressure [14]. However, the variation in oxygen concentration as a function of depth from the exposed surface was not measured and was not correlated to the hardness profile. Gaddam *et al.* [22] collected complementary micro-hardness line profiles and semi-quantitative EPMA oxygen concentration profiles from the surface of Ti-6242, heated in air at 500-700°C. This allowed the extent of oxygen ingress to be measured and related to changes in mechanical properties. However, the oxygen measurements were not fully quantitative, meaning a direct link between absolute oxygen concentration and hardness could not be made. Furthermore, the hardness line profiles sampled limited areas, making it difficult to draw conclusions about the effect of microstructure on oxygen ingress and hardness.

With the advent of fast indentation, it has become possible to produce arrays containing thousands of indents in a matter of hours, allowing hardness to be mapped over large areas of microstructure in a variety of materials [24–28]. This opens up the possibility of correlating hardness and composition changes throughout a complex microstructure [29]. Work by Magazzeni *et al.* [30] lays out in detail the experimental protocol and analysis methods required to correlate maps of hardness, oxygen concentration and crystal orientation for commercially pure titanium, heated in air for 230 hours at 700°C. In this work, we apply the methods of Magazzeni *et al.* [30], using a combination of APT, EPMA, nanoindentation and EBSD to study the interplay between hardness, oxygen concentration, microstructure and crystal orientation within the *oxygen-rich layer* formed on an *in-service* TIMETAL 834 compressor disc. This will improve fundamental understanding of underlying mechanisms that govern formation of *oxygen-rich layers* and their effect on mechanical properties. In turn this will ultimately improve component life predictions, enabling components to safely operate for longer at higher temperatures, decreasing required inspection frequencies and costs associated with replacement components.

## 2. Results

A section of TIMETAL 834 (Ti -5.8 wt.% Al -4%Sn -3.5%Zr -0.7%Nb -0.5%Mo -0.3%Si) compressor disc that has experienced temperatures up to 630°C, for 18,000 hours was provided by Rolls-Royce Plc. After forging of the billet in the α + β phase field, the alloy was solution treated at 1015°C for 2 hours, oil quenched and then aged at 700°C for 2 hours, before being air cooled. This resulted in a duplex microstructure of primary α in a transformed β matrix, which can be seen in Figure 1 (d).

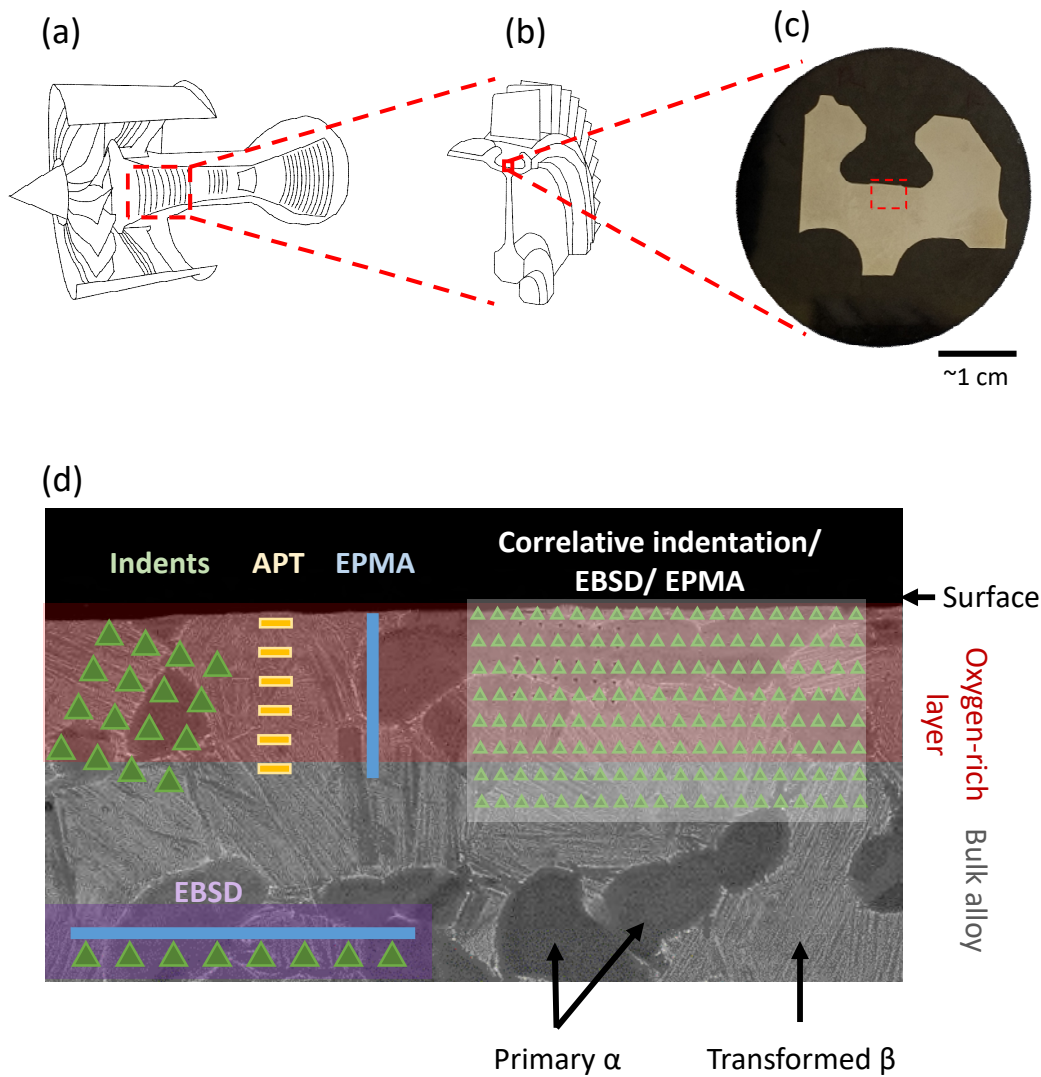

*Figure 1: A cutaway drawing of a jet engine is shown in (a) and in (b) is shown a cutaway drawing of one quarter of a compressor disc. (c) shows a photograph of the cross section of compressor disc used in this study, and (d) shows a schematic of the analysis performed on the TIMETAL 834 in-service compressor disc sample in the current study.*

β stabilisers that partition to the β phase, such as Mo and Nb, have a higher atomic number than typical α phase stabilisers, such as Al. Thus, the β phase can be seen in Figure 1 (d) as white, high contrast regions around the edge of the large α grains, as well as in the α/ β lath regions. A cross-section of the disc mounted in conductive resin, as shown in Figure 1 (c), was ground and polished to a colloidal silica finish. On a separate piece of non-exposed TIMETAL 834, an *oxygen-rich layer* was also created by ageing in air at 550°C for 1000 hours. This sample will be referred to as the *lab-controlled* sample.

## 2.1. Oxygen concentration

Four EPMA profiles, taken to measure the oxygen concentration from the surface inwards on the *in-service* sample, are presented in Figure 2 (a). The location on the surface at which the profiles were taken was chosen at random, yet the shape of the profile was found to be very similar for all four separate measurements. This is not surprising given the uniform temperature distribution experienced by the component.

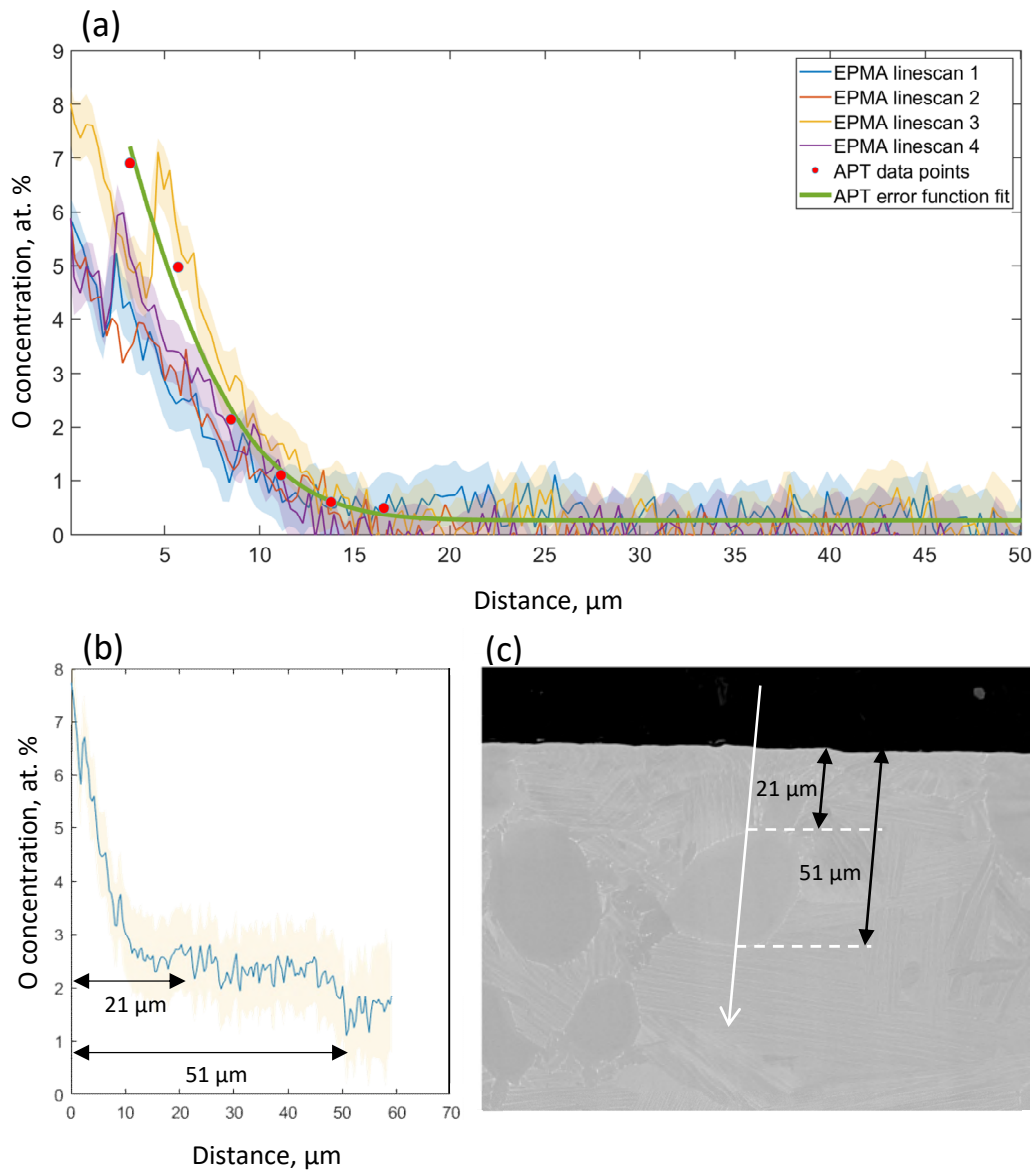

Figure 2: (a) Overlay of the averaged EPMA profiles and the fully quantitative oxygen measurements taken using APT at the surface of the TIMETAL 834 in-service compressor disc sample. Dark blue line shows error function fit to the APT data points. 95% confidence intervals, representing uncertainty in APT-measured oxygen composition, are too small to be visible on the plot. (b) EPMA-measured oxygen concentration profile, calibrated by APT and (c) corresponding SEM image indicating the microstructure over which the profile was taken. Error bars on the EPMA measurements show the standard deviation for the oxygen concentration at each point on the line profile.

It was not possible to quantify the EPMA-measured oxygen without complementary characterisation techniques because of a peak overlap between the K α oxygen and L β titanium x-ray lines in the WDS spectrum. Deconvolution of these peaks requires a standardised oxygen-free Ti sample. A suitable such standard, free from the presence of any surface oxides, could not however be obtained. Consequently, the background-corrected counts have been used, and this means confidence can be placed in the measurement of relative changes in EPMA-measured oxygen content, but absolute values should not be relied upon when the EPMA data is taken in isolation.

To overcome this issue and directly quantify oxygen within the subsurface region, a series of APT samples were made at increasing depth from the compressor disc surface, with the total oxygen content of each APT sample measured. In Figure 2 (a) these APT-measured oxygen concentrations (red circles) have been overlaid with the EPMA profile. A correction shift in the EPMA y-axis (oxygen concentration) was necessary to completely overlay the EPMA and APT profiles, but no shift was required in the x-axis (depth beneath sample surface). Thus, we demonstrate how direct measurement of oxygen with APT can be used to calibrate EPMA profiles, with both techniques combining in a highly-complementary manner which adds confidence to their interpretation.

Figure 2 (a) also shows the complementary error function, which has been fitted to the APT data. Diffusion of oxygen into the surface of the component is a form of non-steady state diffusion and can be described by a specific solution of Fick's second law [32]:

$$C(x,t) = C_s - (C_s - C_0)erf\left(\frac{x}{2\sqrt{Dt}}\right) \qquad \text{(Equation 2)}$$

where $C_s$ is the initial concentration of oxygen at the surface of the material, $C_0$ is the concentration of oxygen in the bulk of the material, $x$ is the diffusion depth, $D$ is the diffusion coefficient and $t$ is time. Equation 2 is reached by applying the following boundary conditions:

1. $C(x=\infty, t) = C_0$
2. $C(x=0, t) = C_s$

The APT data was fitted to the model by varying $C_0$, $C_s$ and $D$ using orthogonal distance regression. The optimised values obtained from this are $C_s$ = 11.5 at. %, $C_o$ = 0.28 at. % and $D$ = 3.1E-07 µm²/ s, respectively.

Although the overall shape and depth of the EPMA profiles is consistent, subtle variations can be seen in some individual profiles. For example, in Figure 2 (b), a slight increase in oxygen concentration is seen at a depth of 21-51 µm. From looking at the corresponding SEM micrograph in Figure 2 (c) of the area profiled, it can be seen that these depths correspond to a large, equiaxed α grain, indicating clear dependence of oxygen concentration on microstructure.

### 2.1.1. Lab Controlled sample

Since the environmental conditions that the compressor disc was exposed to cannot be well defined, a piece of TIMETAL 834 was heated in air for 1000 hours at 550°C to form an *oxygen-rich layer* under controlled conditions, referred to as the *lab-controlled* sample. APT samples were taken at increasing depth from the surface of the sample to measure the oxygen concentration profile from the surface inwards. The EDS maps, EBSD pattern quality map and IPF X map in Figures 3 (b)-(i), (j) and (k), respectively, show the microstructure and elemental segregation within the region from which the APT samples were made. It can be seen from Figure 3 (a) that the section of the APT liftout bar nearest the sample surface (white box) contains an α/ β lath region. Figures 3 (e), (f) and (i) show segregation of Nb, Mo and Zr to the β phase in this lath region. The two grains which make up the majority of the APT liftout bar can be seen in Figure 3 (k) to be of similar crystallographic orientation, despite having different microstructures.

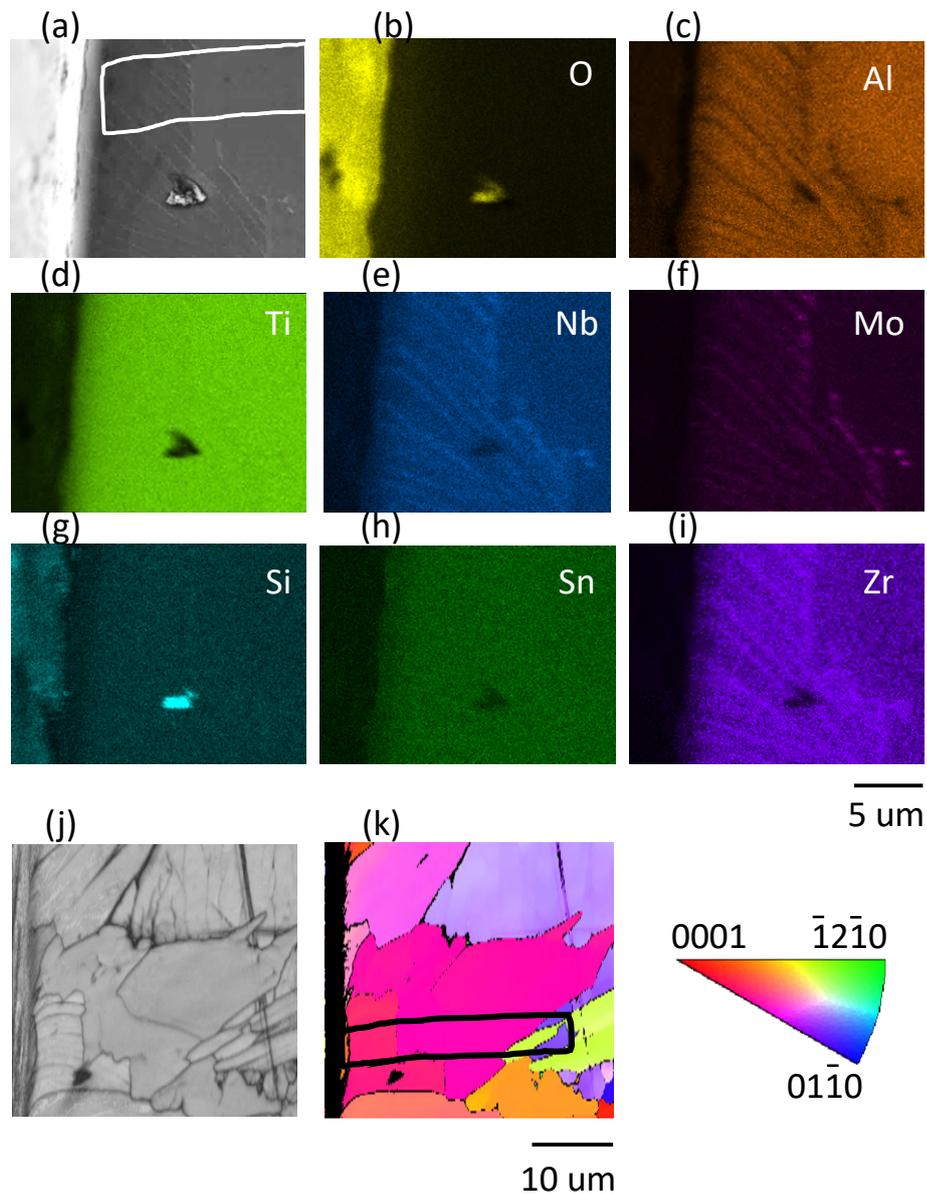

*Figure 3: (a) SEM image and corresponding EDS maps for (b) O, (c) Al (d) Ti, (e) Nb, (f) Mo, (g) Si, (h) Sn and (i) Zr showing distribution of elements within the region from which APT samples were prepared in the TIMETAL 834 lab-controlled sample. (j) EBSD pattern quality map and (k) IPF X image the microstructure and the grain orientation of the region from which the APT liftout was created. The rectangular boxes drawn on (a) and (k) indicate the position of the APT liftout bar. The Si and O rich particle also visible in the EDS maps is a piece of surface contamination.*

Figure 4 shows the fitted oxygen concentration profile for both the *in-service* and *lab-controlled* sample, for comparison. The optimised diffusion values obtained for the *lab-controlled* sample are $C_s$ = 11 at. %, $C_0$ = 0.35 at. % and $D$ = 1.4E-06 µm²/ s, respectively. The activation energy for formation of the *oxygen-rich layer* can be calculated from this value of diffusion coefficient using

$$D = D_0 e^{-\frac{Q}{RT}} \quad \text{(Equation 3)}$$

where $D_0$ is the diffusivity, $Q$ is the activation energy for *oxygen-rich layer* formation, $R$ is the gas constant and $T$ is temperature. A literature $D_0$ value for TIMETAL 834 of 10 mm²/s [33] was used to calculate a value of $Q$ of 203 kJ/ mol. This is higher than reported values of activation energy for *α case* formation in α + β TIMETAL 834, (160-184 kJ/ mol) [33–35] but is of the same order of magnitude. The activation energy for *α case* formation in β solution treated TIMETAL 834 is reported to be 223 kJ/ mol [34].

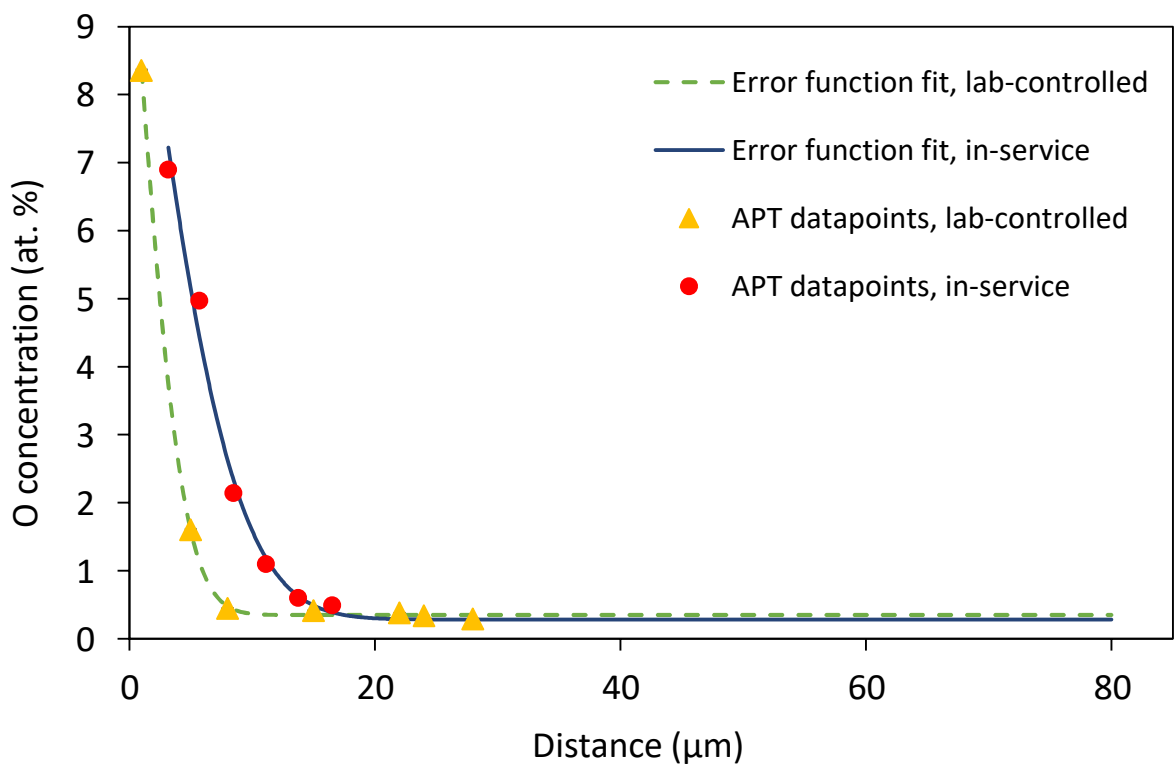

Figure 4: APT-measured oxygen concentration profiles and fitted error function in the in-service component and lab-controlled sample. 95% confidence intervals, representing uncertainty in APT-measured oxygen composition, are too small to be visible on the plot.

Thus it seems activation energy is sensitive to β volume fraction, and it is possible that the β volume fraction in the material in this study is higher than usual.

Taking $Q$ to be the value calculated for the *lab-controlled* sample, and $D_o$ to be 10 mm$^2$/s [33], we can use Equation 3 to calculate a representative operating temperature experienced by the component. The temperature calculated is 510°C, which is within the operating temperature window of the component.

Figure 5 (a)-(e) shows APT 3-D atom maps from the *lab-controlled* sample for the specimen taken from nearest the surface, which has the highest oxygen concentration. Within the APT reconstruction, a Mo-rich phase can be seen. Adjacent to this phase is a region enriched with Sn, Nb and TiO, and depleted in Al and Zr, as shown by the isoconcentration surfaces in Figure 5. While the isoconcentration surfaces are used qualitatively to visually demarcate the phases, for quantitative analysis, the Al isoconcentration surface indicated in Figure 5 (a) was used to create overlap-solved proximity histograms [36], seen in Figure 5 (f). The Mo-rich phase was isolated using a 2.0 at. % isoconcentration surface and found to be of composition (at. %) Ti0.64-Mo0.23-Al0.02-Nb0.07.

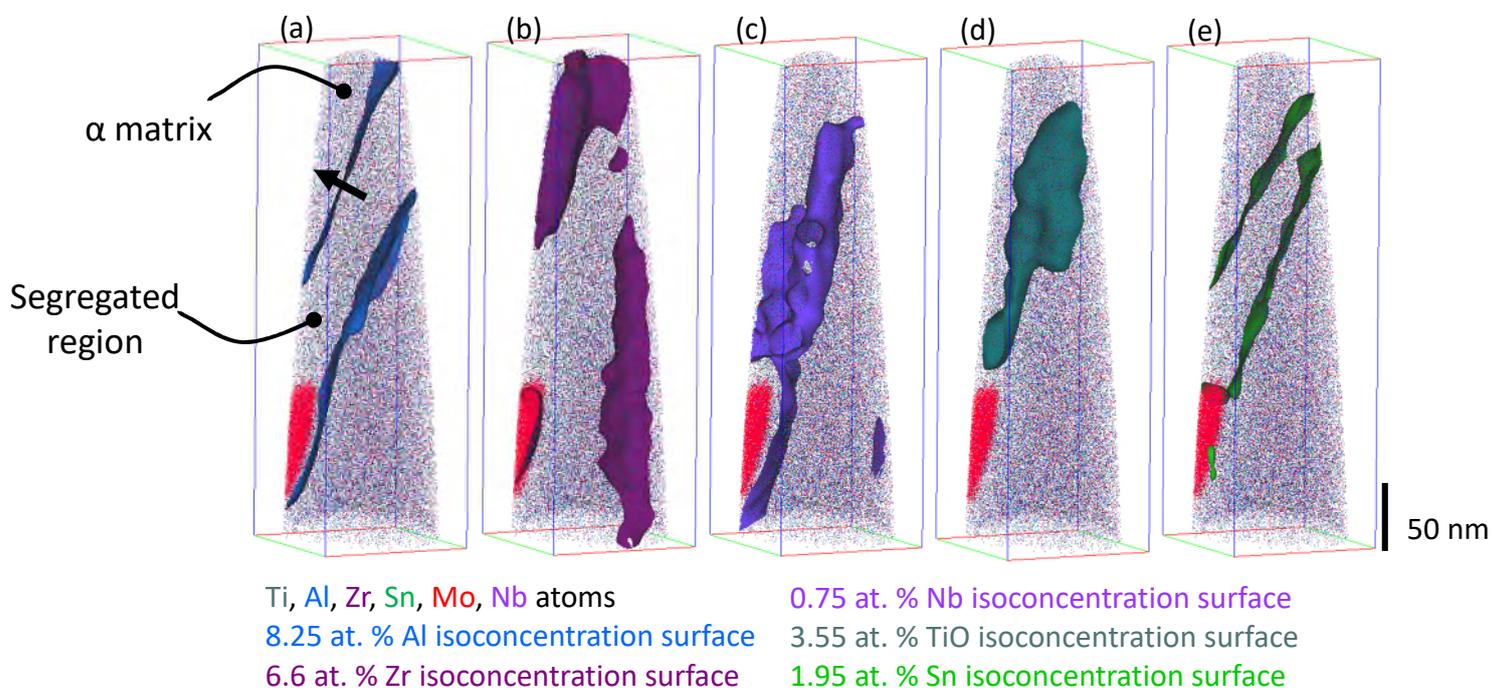

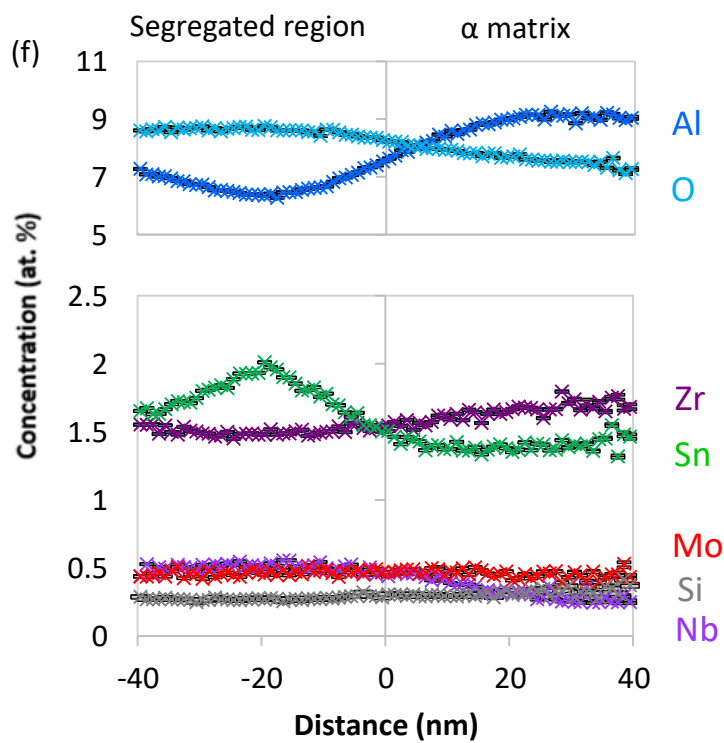

Figure 5: APT atom maps for the APT sample nearest the surface of the TIMETAL 834 lab-controlled sample. Isoconcentration surfaces show the segregation of (a) Al and (b) Zr away from a possible α-β boundary and segregation of (c), Nb, (d) TiO and (e) Sn along a possible α / β phase boundary. (f) Overlap-solved proximity histograms (proxigrams) calculated across the Al isoconcentration surface indicated by the black arrow in (a). The proxigrams quantify the enrichment of O, Sn and Nb, and the depletion of Al and Zr, in the segregated region. Error bars show 95% confidence intervals, representing the uncertainty in composition due to counting error and solving mass spectrum peak overlaps.

## 2.2. Nanohardness measurements

Two indent arrays, taken normal to the oxygen-exposed surface, give rise to the hardness profiles at two separate locations labelled *site A* (black circles) and *site B* (red squares) shown in Figure 6. The hardness profile from *site A* is the same shape as that at *site B*, but the hardness at *site A* is consistently slightly higher.

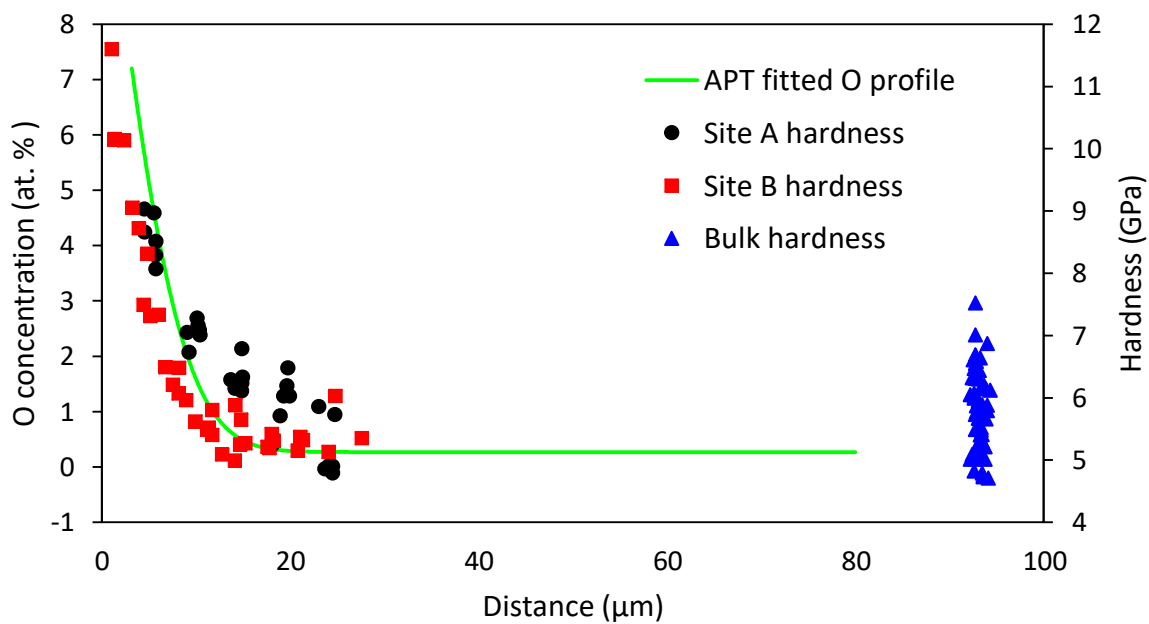

*Figure 6: Overlay of the APT fitted oxygen profile (green) and the two surface hardness arrays (black and red), measured on the TIMETAL 834 in-service compressor disc. In blue are the hardness measurements taken from the bulk indents at a constant depth of 95 um from the surface.*

### 2.2.1. Bulk material

To explore the cause of this hardness difference, and to investigate if it was related to microstructure and crystal orientation, a set of indents were made in the bulk of the compressor disc material. Indents were made at a constant depth of 95 μm from the surface, away from the influence of the oxygen profile. The hardness values from this can be seen in Figure 6 (blue triangles), while the results are presented in more depth in Figure 7 (a). The microstructure of the region indented can be seen clearly in the backscattered electron image in Figure 7 (b). The corresponding EPMA line scan in Figure 7 (c) shows that oxygen concentration measured adjacent to the bulk indents appears relatively uniform.

Grain orientation information was obtained from the IPF Z map of the indented region in Figure 7 (d), and the second Euler angle was extracted for each indented grain, which is equal to the declination angle in the hexagonal α-Ti crystal system [37].

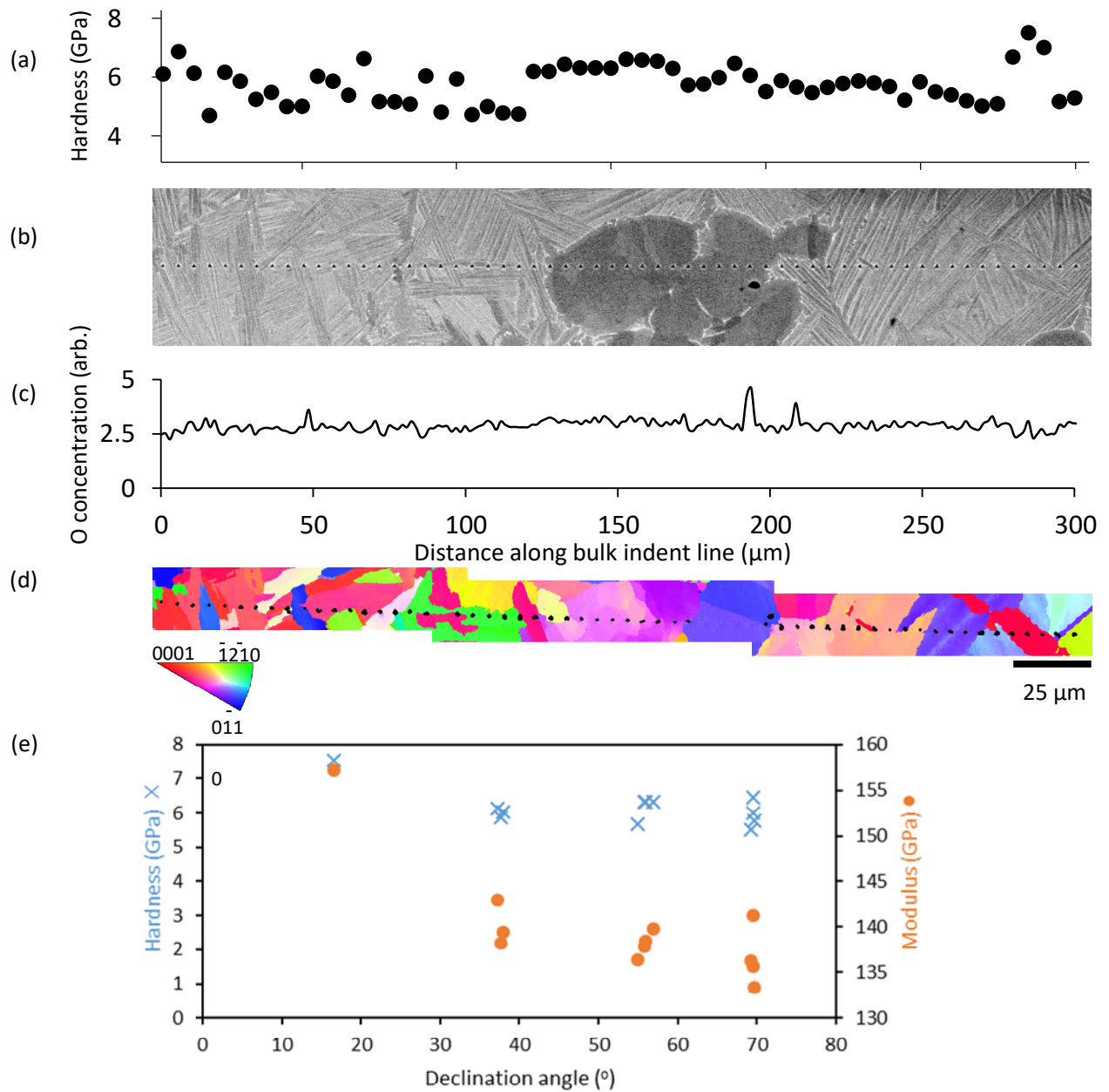

Figure 7: (a) Hardness plotted along the length of the indent profile in the bulk of the TIMETAL 834 in-service compressor disc. (b) BSE image showing the microstructure of the region indented. Contrast between the large α grain in the middle of the image and the surrounding α / β laths can be seen. (c) EPMA-measured oxygen line profile measured adjacent to the indent profile, showing the oxygen concentration is relatively constant. Distance is measured in microns along the length of the indent profile, at a constant depth of 95 μm from the exposed surface. (d) IPF Z map showing the orientation of the indented grains in the bulk material at a constant depth of 95 μm from the surface. (e) Hardness and modulus of the bulk TIMETAL 834 compressor disc sample, plotted as a function of grain declination angle.

The declination angle is the angle between the crystal c-axis and the indent loading direction and was combined with the hardness measurement for each indent that did not interact with a grain/ phase boundary. The results of this are presented in Figure 7 (e), which shows that grains which have a small angle between their c-axis and the sample surface normal are harder, in good agreement with trends measured in the literature [38,39].

### 2.2.2. *Oxygen-rich layer*

To further explore the interrelationship between microstructure, crystal orientation, oxygen concentration and hardness throughout the *oxygen-rich layer*, correlative hardness maps, EPMA maps and EBSD maps were created at the surface of the compressor disc sample, as shown schematically on the right-hand side of Figure 1 (d).

Figure 8 (a) shows the region of interest at the surface of the compressor disc. In Figure 8 (b), the oxygen EPMA map reveals some correlation of oxygen content with microstructure. Oxygen counts are displayed on the map, which can be taken to be indicative of oxygen concentration. Some grains and lath features can be seen on the hardness map in Figure 8 (c). Figure 8 (d) shows the change in grain orientation between different regions of microstructure. Comparing Figure 8 (a) and Figure 8 (d) reveals that some of the large α grains contain multiple regions of different orientations. These independent maps of oxygen content, hardness and grain orientation are all useful separately, but overlaying the data enables trends and relationships to be analysed in much greater depth. Figure 8 (c) was created by tiling eight hardness arrays. The vertical lines are from where two adjacent hardness arrays have overlapped slightly, due to the limited precision in the stage movement of the nanoindenter.

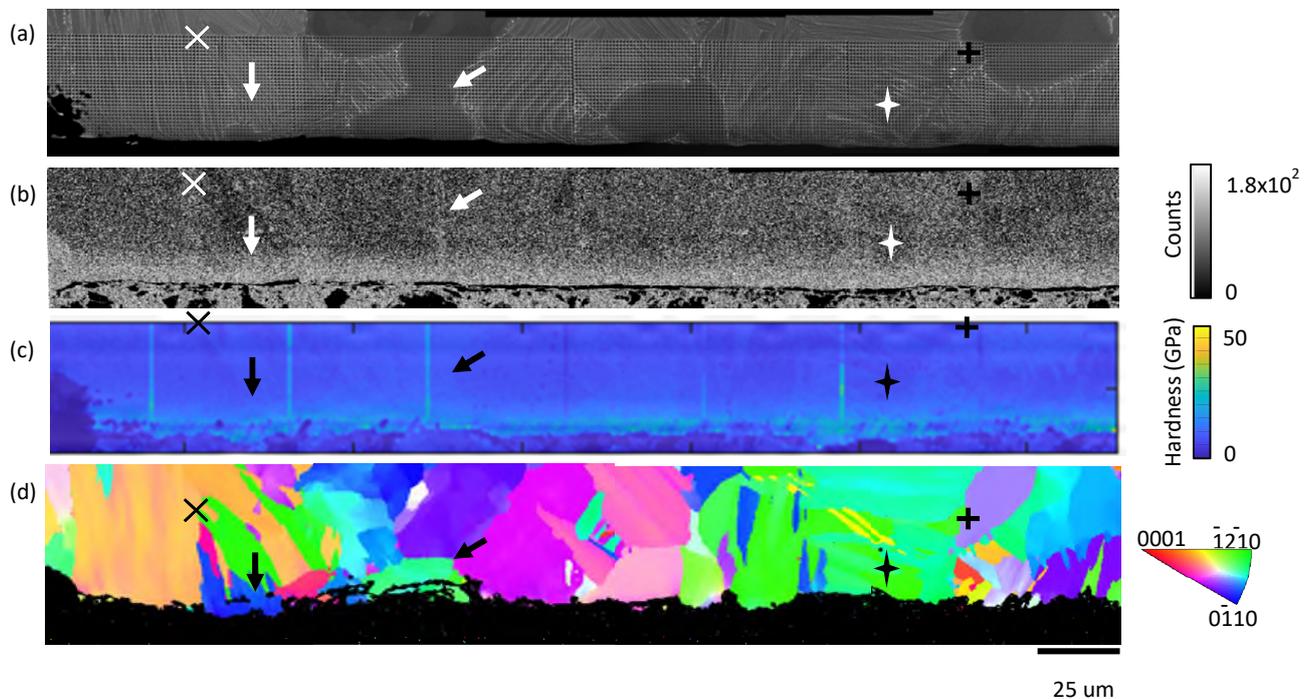

Figure 8: (a) shows the backscattered electron image of the indented region of interest at the surface of the in-service TIMETAL 834 compressor disc. (b) shows the electron probe microanalysis oxygen map (c) shows the hardness map and (d) shows the IPF Z map, all of the same region of interest. The five sets of symbols indicate examples of common features on each of the maps.

In Figure 9 (a), hardness is plotted as a function of the second Euler angle, phi, and oxygen content, for both the primary α grains and lath regions of microstructure. Clearer trends can be seen in the plot for the extracted α grain data in Figure 9 (b). The hardness was plotted as a function of declination angle for the α grains in Figure 9 (c), and for the laths in Figure 9 (d). A linear fit and 95% confidence bounds have been applied. In Figure 9 (e), the hardness vs declination angle for the α grains is plotted for just the low oxygen content data, as indicated by the dotted red lines on Figure 9 (b). The trends on Figure 9 (c)-(e) are marginal overall, with considerable scatter, but it can still be seen that only the gradient of the fit of Figure 9 (e) is negative. The hardness of α-Ti is expected to increase with decreasing declination angle, in line with the trend observed for the bulk TIMETAL 834 material in Figure 7 (e). However, this trend is only seen in Figure 9 (e) suggesting that, at higher oxygen contents and when β phase is present, factors other than α grain orientation have a greater effect on hardness.

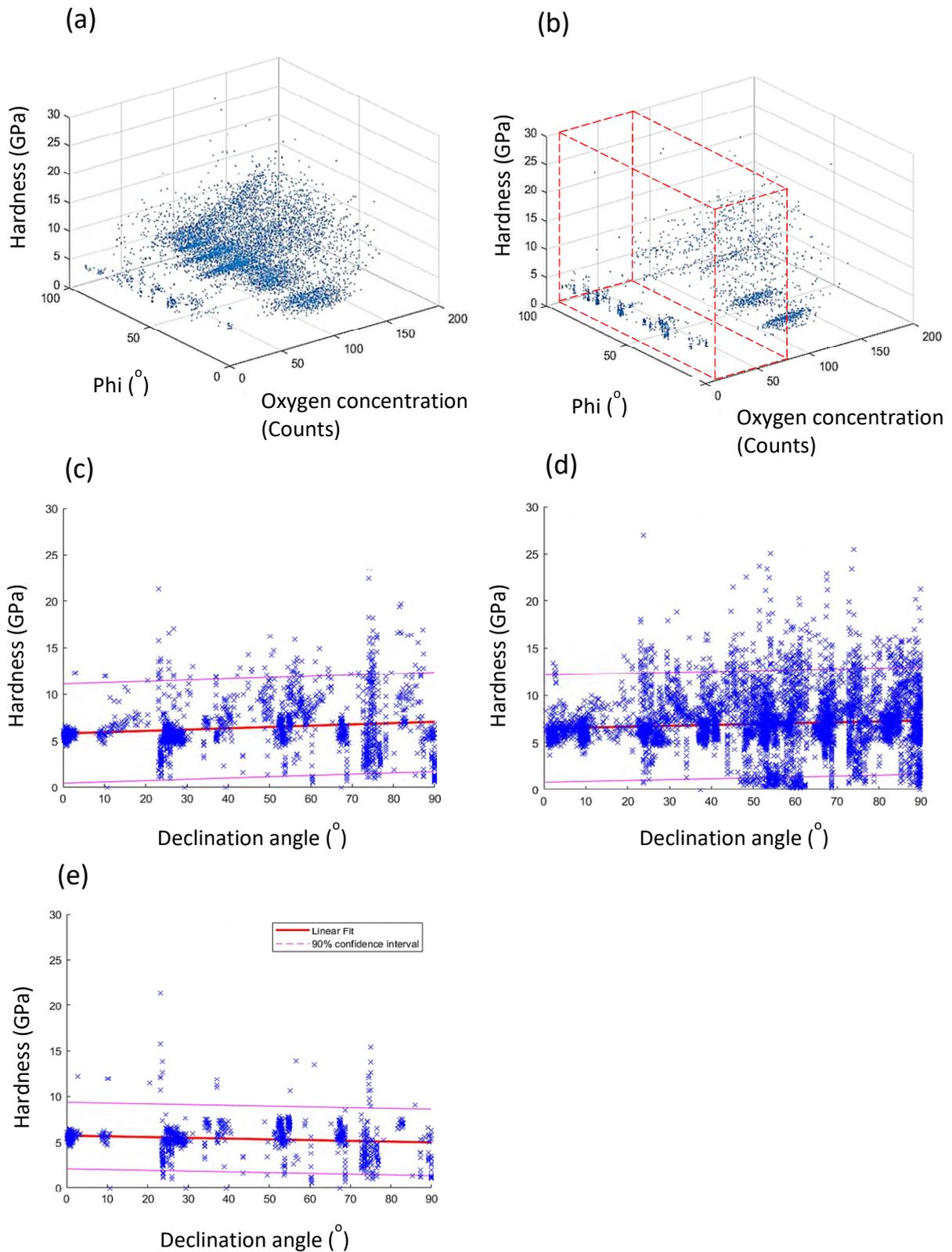

*Figure 9: Three-dimensional plots showing the relationship between hardness (collected through nanoindentation mapping), oxygen concentration and grain orientation for (a) all regions of microstructure and (b) primary α grains only. Plots of hardness as a function of declination angle, with 95% confidence intervals and a linear trend line for the (c) α grains, (d) lath regions, (e) α grain regions with low oxygen. The α grain regions with low oxygen that are used to plot (e) are indicated by the dotted red box drawn in (b).*

## 3. Discussion

### 3.1. Oxygen ingress

Excellent agreement between the shapes of the EPMA- and APT-measured oxygen profiles, as seen in Figure 2 (a), validates the use of APT to produce fully quantitative oxygen profiles from a set of discrete measurements. This is of particular importance since accurate quantitative oxygen measurements in titanium are challenging to obtain with other techniques on this length scale.

Figure 2 (b) and (c) show that oxygen concentration within the *oxygen-rich layer* has some dependence on microstructure; there is a clear drop in oxygen concentration from the large equiaxed α grain to the fine α/ β lamellar structure underneath. A number of factors are known to affect oxygen diffusivity, and thus penetration depth, in near-α titanium alloys. Leyens *et al.* [40] observed that the depth of the oxygen diffusion zone is greater for bimodal microstructures compared to lamellar microstructures. They attribute this to the larger number of phase boundaries available in the bimodal microstructure, which can act as fast diffusion paths for oxygen. However, more recent work by Satko *et al.* [4] shows the opposite trend; the *oxygen-rich layer* is actually deeper for the lamellar microstructure. This can be explained by considering the dependence of oxygen penetration on the alignment of interfaces relative to the oxygen-exposed surface; α/ β phase boundaries perpendicular to the sample surface increase oxygen ingress [41] and thus it is possible that a large proportion of boundaries in the Leyens lamellar microstructure were unfavourably oriented. The α/ β laths across which oxygen concentration was measured in Figure 2 (c) are oriented almost parallel to the oxygen-exposed surface. This unfavourable orientation of the α/ β phase boundaries results in lower oxygen diffusivity, which could account for the lower oxygen concentration in the α/ β lath region compared to the equiaxed α grain.

The extent of oxygen penetration for each of the *in-service* and *lab-controlled* samples was determined by calculating the depth at which the gradient of the fitted oxygen concentration profiles, in Figure 4, dropped to below 1 ppm O/micrometre. The oxygen penetration depths were found to be 27.6 μm and 14.3 μm for the *in-service* and *lab-controlled* sample, respectively. The deeper oxygen penetration depth for the *in-service* sample is expected since the exposure time for an *in-service* component is typically much longer than 1000 hours, and the flight cycles experienced are more complex than the isothermal treatment given to the *lab-controlled* sample. Lack of microstructural change (β denudation) observed in the *lab-controlled* and *in-service* material also points towards both samples having experienced similar temperatures below the β transus locally.

The APT dataset taken from nearest the surface of the *lab-controlled* sample, seen in Figure 5, contains a small volume of a Mo-rich phase. The composition measured, Ti0.64-Mo0.23-Al0.02-Nb0.07, is consistent with that of the ordered intermetallic $Ti_2$(Al, X) phase, where X= Nb, Mo [42].

The isoconcentration surfaces in Figure 5 (a)-(e) shows Sn and Nb segregate in a plane ahead of the Mo-rich phase, and Al segregates strongly away from this plane. This is supported by the proximity histograms in Figure 5 (f). The EDS maps in Figure 3 show Nb partitioning to the β phase, while Al is a strong α stabiliser and therefore partitions away from β phase. Thus, it is possible that the plane ahead of the Mo-rich phase is a piece of β phase that is depleted in β stabilising elements as a result of formation of the Mo-rich phase. When examined in context of the SEM and EDS images in Figure 3, it can be seen that the region nearest the surface from which this APT sample was taken contains α/ β laths, which are oriented at an acute angle to the oxygen-exposed surface, and thus favourably oriented to act as fast diffusion paths for oxygen. Inspection of the APT maps corresponding to this surface sample in Figure 5 does indeed reveal segregation of TiO ahead of the Mo-rich phase. Therefore, Figure 5 (c) is actually showing oxygen diffusing preferentially along an α/ β boundary that is believed to be acting as a fast diffusion path, ahead of a Mo-rich phase.

### 3.2. Contributions to hardness

Hardness has been directly measured as a function of depth from the surface of the *in-service* compressor disc and found to independently agree with the shape and depth of the oxygen profiles, as can be seen in Figure 6. This supports the idea that interstitial oxygen alone is causing the hardening, and that there is no *β denudation*, in line with the definition of *oxygen-rich layer* introduced by Satko *et al.* [4].

Analysis of bulk indents, away from the oxygen-exposed surface of the compressor disc, shows that grains with a small declination angle are harder, and thus the crystal orientation has an effect on the measured hardness. The oxygen EPMA profile in Figure 7 (c) shows no significant change in oxygen concentration along the length of the bulk indents, meaning oxygen can be discounted as a variable affecting relative hardness changes in the bulk material, at least at the same depth within it. However, it is likely that microstructure is playing a role in the observed variation in bulk hardness, seen in Figure 7 (a). Ghamarian *et al.* [43] report that equiaxed α grains are the softest microstructural components in α + β alloys, such as TIMETAL 834. When Figure 7 (a) and (b) are compared, lower hardness regions correspond to areas where there are large α grains. The relatively small area of microstructure sampled in Figure 7 makes it difficult to distinguish the contributions of microstructure and grain orientation to bulk hardness. However, we can conclude that both microstructure and grain orientation are giving rise to the variation in the bulk hardness measurements in Figure 7 (a). This likewise also leads to the suggestion that grain orientation and microstructure are contributing to the variation in the height of the surface hardness profiles at *sites A* and *B*, in Figure 6.

### 3.2.1. Correlative hardness, EBSD and EPMA mapping

For the oxygen-exposed *in-service* compressor disc, changes in hardness due to crystal orientation and microstructure, as outlined in Figures 6 and 7, are minor compared to the change in hardness as a function of depth due to oxygen ingress from the surface. Correlative nanoindentation, EPMA, EBSD and BSE imaging was used to study the interdependence of crystal orientation, microstructure, oxygen ingress and hardness of the *oxygen-rich layer* in greater detail, enabling consideration of the various possible mechanisms contributing to changing mechanical properties.

The dependence of hardness on crystal orientation for HCP α-Ti stems from the difference in critical resolved shear stress on each of the slip systems involved in the deformation. The critical resolved shear stress is higher on the <c+a> type slip systems, meaning the material is harder when indented closer to the c-axis [38,44].

The cubic β phase does not have the highly anisotropic properties of the hexagonal α phase. Therefore, any change in hardness as a function of the second Euler angle (equal to the declination angle for the α phase) is expected to be less pronounced for the β phase compared to the α phase. Despite the small volume fraction of β phase present (~ 5 % by volume [3]), there is a high concentration of α/ β boundaries in the lath regions. Therefore, when indenting lath regions, any trend in hardness as a function of crystal orientation is made less clear by the contributions of the cubic β phase and the inevitably large number of lath boundary interactions.

Very shallow (200 nm deep) indents were used to create the hardness map. For a Berkovich indentor, a minimum indent spacing of 10 times the indent depth has been found to be necessary to minimise the influence of neighbouring indents [45]. However, the indents were spaced 1 μm apart to obtain sufficient spatial resolution on the hardness maps. Hence, maps cannot be used to report absolute values of hardness, but comparisons of relative hardness changes remain valid. The size of the plastic zone for each indent is such that multiple laths will be indented at one time, which may contribute to a lack of contrast in nanoindentation hardness.

Nevertheless, we can mitigate the grain boundary and β phase effects described above by extracting the data corresponding to the lath regions and the α grains and analysing them separately. Comparison of Figure 9 (a) and Figure 9 (b) shows that extracting the data for the α grains increases contrast in the hardness measurements. This is further emphasised by the clearer trends in hardness as a function of declination angles that are observed for the extracted primary α grains in Figure 9 (c) compared to the data for the extracted lath regions in Figure 9 (d). Thus it appears the β phase and presence of high density of grain boundaries in the lath regions are altering the observed trend in declination angle.

Regions containing α + β laths are typically harder than α grains [43]. This may be partly because mobile dislocations can travel further in the α grains without being impeded by grain/ phase boundaries. However, lath regions are made up of a series of α colonies, which are approximately the same size as the primary α grains (20-50 μm), as can be seen in Figure

8 (a). Within a colony, all the laths are oriented in the same direction, meaning that if slip occurs in one lath, it is easily transmitted to neighbouring laths of the same orientation. Thus colony size is a more important factor in governing slip length and, by extension, hardness in bimodal microstructures [46]. Figure 8 (a) shows that difference in size of primary α grains and colonies is small, which could explain why there are only small differences in hardness between laths and non-laths, illustrated in Figure 9 (c) and Figure 9 (d).

It is the interaction of oxygen atoms with dislocation cores that produce interstitial hardening. Oxygen in titanium provides exceptional levels of solid solution hardening, as described in detail by Yu *et al.* [47]. The strengthening originates from a strong interaction between the oxygen atom and screw dislocation cores. The strong interaction is caused by a combination of strong repulsion of the dislocation by the oxygen solute, a high energy barrier to rearrangement of oxygen atoms within the core and cross slip induced by the oxygen atoms [47].

Comparison of Figure 9 (c) and Figure 9 (e) shows that oxygen is affecting the observed trend in hardness as a function of declination angle. In α-type titanium, hardness is expected to increase as the declination angle decreases [38]. However, this is only observed in the α grain regions when the high oxygen contributions (greater than 80 arbitrary units) are filtered out. At higher oxygen concentrations, it seems that the effect of interstitial oxygen solid solution hardening is dominant. Thus, the effects of microstructure and crystal orientation are only significant when the oxygen content is low, i.e. further away from the surface.

Although crystal orientation can make slip challenging to initiate, the barrier to dislocation motion presented by oxygen is very large [47]. Likewise, the type of microstructure (lath region or α grain) has minimal effect on hardness since the lath regions and α grains are of comparable size, leading to similar slip lengths. The unusually strong, local dislocation core-interstitial interaction [47] dominates over the long-range, weaker, elastic interaction of dislocations with grain boundaries/ phase boundaries. This explains why interstitial hardening dominates except at low oxygen concentrations.

This work has defined an analysis procedure for correlative hardness mapping, EPMA mapping and EBSD mapping, and highlights factors that need to be taken into consideration for future work. For example, to study the effect of microstructure on oxygen ingress, a larger region of microstructure should be sampled, with multiple α grains of varying declination angle and multiple lath regions at the surface. This would also enable a more quantitative study of the effect of lath orientation relative to the ingress direction on oxygen ingress.

## 4. Conclusions

A detailed, multi-technique approach was taken to characterise the *oxygen-rich layer* on a Ti-alloy compressor disc that has experienced *in-service* civil aerospace gas-turbine engine conditions. Oxygen concentration was directly measured and quantified as a function of depth from the surface using Atom Probe Tomography, and the resultant oxygen diffusion profile obeys Fick's second law. Fitting the APT-measured oxygen profile using the complementary error function yields an activation energy for formation of the *oxygen-rich layer* of 203 kJ/ mol, which is consistent with literature values [33–35].

The oxygen ingress depth was found to be 27.6 µm and 14.3 µm for the *in-service* and *lab-controlled* sample, respectively. Since the conditions experienced by both these samples was equivalent to the intermediate *in-service* temperatures described by Satko *et al.* [4], no β denudation was observed and the oxygen ingress depth is considered to be the depth to which the mechanical properties are detrimentally affected.

The shape of the APT- and EPMA-measured oxygen profiles agree, as do the surface hardness profiles, indicating that interstitial oxygen is the main hardening mechanism within the *oxygen-rich layer*. The contributions of microstructure and crystal orientation to the hardness measurements have been studied. Correlative hardness mapping, EBSD mapping and EPMA mapping has been used to find the relative contributions of oxygen concentration, crystal orientation and microstructure to hardness throughout the oxygen-rich layer. Oxygen content as a function of depth has a greater impact on hardness than microstructure and crystal orientation, highlighting the importance of using APT to accurately quantify oxygen concentration. Moreover, oxygen ingress has been shown to be affected by microstructure on both the micron-scale, using EPMA, and on the nanoscale, through use of APT. Both techniques underline the effect of α/ β interfaces on oxygen ingress. In addition to improving lifetime prediction models, this study suggests that control over the proportion and orientation of α/β interfaces at the surface offers the potential to both reduce oxygen ingress and significantly increase in-service lifetimes for TIMETAL compressor blades.

# 5. Methods

## 5.1. EPMA line profiles

EPMA background-corrected X-ray line profiles were acquired on five randomly selected points on the surface of the compressor disc. The O Kα, Al Kα, Zr Lα, Sn Lα and N Kα signals were collected using a CAMECA SX5-FE system in the Department of Earth Sciences, University of Oxford using a 15 keV and 10 nA focused electron beam. A relatively long dwell time of 130 s per pixel was employed to improve the counting statistics.

## 5.2. Surface nanoindent arrays

A NanoIndenterXP (MTS) was used to create two indent arrays at the surface of the compressor disc, one of 10 by 5 indents, and one of 6 by 6 indents. Indents were 250 nm deep to allow for close spacing (4 µm) such that multiple measurements could be made within the alpha-case region. The shallow indent depth also improved sensitivity to the fine microstructural features within TIMETAL 834. Indents were made using a diamond Berkovich indenter. The contact stiffness was continually measured using the continuous stiffness measurement (CSM) technique (2 nm amplitude, 50 Hz oscillation) [48] and was used to assess the quality of the indentation data. All indents used here have a near-linear depth vs load/stiffness, which indicates suitable sample mounting and good contact between tip and surface. Consequently, the error bars on individual hardness measurements are so small that they lie within the data point and cannot be seen on a plot. Hardness values reported are derived from the CSM data and averaged between depths of 100-250 nm.

### 5.3. EBSD and EDS

Electron Backscatter Diffraction (EBSD) and EDS was carried out using a Zeiss Crossbeam 540 Focussed Ion Beam/Scanning Electron Microscope (FIB/SEM) instrument equipped with Oxford Instruments XmaxN 150 EDS detector and Oxford Instruments Nordlys Max EBSD detector. A probe current of 15 nA, an aperture size of 120 mm and an accelerating voltage of 30 kV were used for EBSD, whilst an accelerating voltage of 20 kV was used for EDS. Multiple EBSD maps were taken of the indents in the bulk compressor disc material, with step sizes of 0.5 µm, and the maps were subsequently stitched together. An interaction volume for each indent was defined as 4 µm. For each indent that did not interact with a grain/phase boundary, the orientation of the indented grain was determined.

### 5.4. APT

APT samples were prepared via the standard FIB lift-out method [49]. For both the *lab-controlled* and *in-service* material, APT samples were made at increasing depth from the surface, as shown schematically in Figure 1 (d), and then analysed with a CAMECA LEAP 5000 HR system, using a laser energy of 40 pJ, a stage temperature of 50 K and a pulse frequency of 200 kHz. APT data analysis was performed using the commercially available Integrated Visualisation and Analysis Software (IVAS 3.8.2) software package (CAMECA) and the open-access 3Depict software [50]. The oxygen content of each APT sample was analysed by solving the mass spectrum peak overlaps using custom Matlab code [51]. 95% confidence intervals are reported as error bars on APT composition measurements and are a measure of the uncertainty in the composition, which was determined using the maximum likelihood estimation when mass spectrum peak overlaps were present. Further details of the sources of uncertainty in APT measurements can be found in work by London [52].

### 5.5. Correlative hardness, EBSD and EPMA mapping

EPMA oxygen maps were acquired on the region of interest at the surface of the compressor disc, prior to indentation, using a CAMECA SX5-FE electron microprobe analyser (EPMA) in the Department of Earth Sciences, University of Oxford. A 10 keV and a focused 10 nA electron beam was employed, with a dwell time of 0.05 s per pixel. To minimise carbon deposition during analyses, an in-chamber liquid nitrogen cold plate was employed.

EBSD was then carried out on the same region of interest, using a Zeiss Crossbeam 540 Focussed Ion Beam/Scanning Electron Microscope (FIB/SEM) instrument equipped with Oxford Instruments Nordlys Max EBSD detector. A beam current of 15 nA, an aperture size of 120 mm and an accelerating voltage of 30 kV were used for EBSD, with a step size of 0.3 µm. EPMA was carried out prior to EBSD in order to avoid any potential surface carbon contamination arising from the EBSD compromising the oxygen analyses.

An express indent array of area 350 µm by 50 µm, corresponding to tens of thousands of indents, was created to produce hardness maps for the region of interest at the surface of the exposed compressor disc. Indents were performed on the Agilent Nano indentor G200 using a diamond Berkovich tip and 200 nm deep indents at a spacing of 1 µm and at a load of 4 mN.

The array of hardness measurements was used as the reference grid to which all other map data was aligned. Pairs of control points, distinct features that were identifiable on both the BSE image and the hardness map, were manually selected and used to infer a geometric transform that would overlay the two sets of data. Features used as control points include grain boundaries and phase interfaces, and five sets of control points are represented with symbols on Figure 8, as an example. The EPMA and EBSD data was overlaid with the hardness data in the same way.  The BSE image was used to segment the area studied into primary α grains and α/β lath regions by making use of the lower contrast of the α grains compared to the lath regions. This enabled hardness, declination angle of the α phase (mined from the EBSD data) and oxygen content measurements to be extracted separately for the lath regions and α grains. For example, the grain orientation maps corresponding to the α grain and lath regions can be seen in Supplementary Figure 1. A hardness threshold was selected such that data below 4.5 GPa, mainly corresponding to the indents in the Bakelite mounting material, was discarded. Supplementary Figure 2 shows the hardness data that was retained.

The code used for analysing the indentation map data and correlating with EPMA and EBSD data was written in MATLAB R2019a and is available in the following Github repository under an MIT License: https://github.com/cmmagazz/XPCorrelate.

## Acknowledgements

EPSRC and Rolls-Royce plc supported the work in this paper under an iCase agreement, under project EP/N509711/1. The EPSRC funded the UK National Atom Probe facility in Oxford under project EP/M022803/1. The authors are grateful for the advice and services offered at the David Cockayne Centre for Electron Microscopy in the University of Oxford.

# Supplementary information

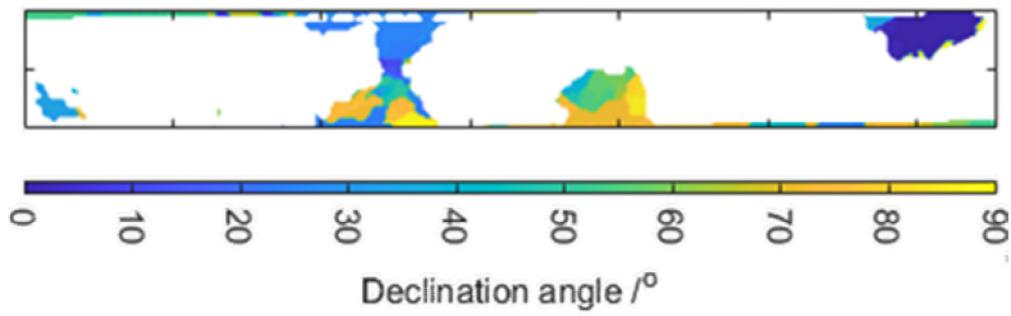

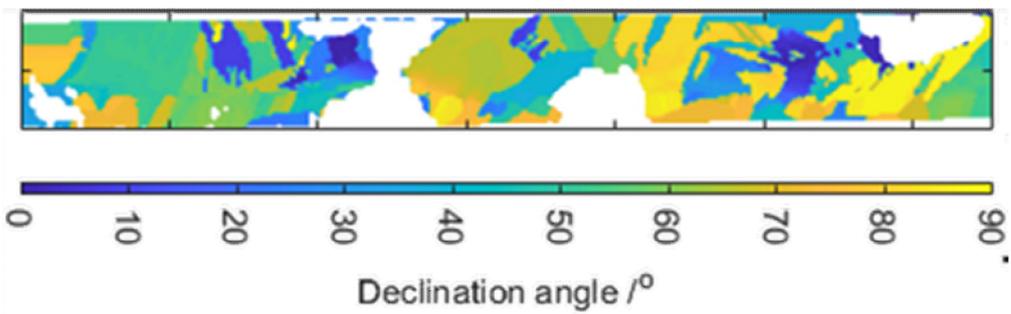

Supplementary Figure 1: Map of the phi angle showing the extracted regions of microstructure corresponding to the (a) α grains and (b) lath regions.

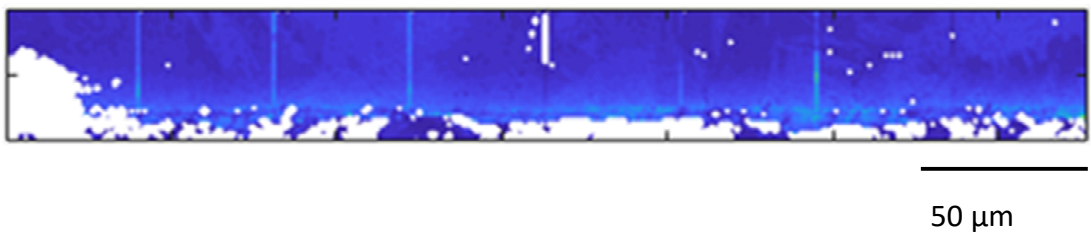

50 µm

Supplementary Figure 2: Hardness map showing the data retained after filtering out hardness values below 4.5 GPa.